# The simulator of the VLT Deformable Secondary Mirror: a test tool for adaptive optics instruments for the Yepun-UT4 telescope

Runa Briguglio*[a], Armando Riccardi, Luca Carbonaro[a], Enrico Pinna, Chiara Selmi[a], Paolo Grani[a]

[a]INAF Istituto Nazionale di Astrofisica, Osservatorio Astrofisico di Arcetri

## ABSTRACT

The Deformable Mirror Simulator (DMS) is an optical device reproducing the F/13 beam from the adaptive secondary mirror of the Very Large Telescope UT4. The system has been designed and integrated as a test tool for the calibration and functional verification of the WaveFront sensor module of the ERIS instrument (or ERIS-AO). To this purpose the DSMSim includes a high order deformable mirror and two sources to mimic the laser and natural asterisms and illuminate the WFS optics.

In this paper we report the design of the DSMSim, the integration, verification and alignment procedure with the ERIS-AO; in the end we outline a roadmap for future improvements of the system. This work is intended to be a reference for future intrumentation projects (e.g. MAVIS-AO) for the VLT.

## 1 CONCEPT AND OPTICAL DESIGN

The ERIS[1] instrument (Enhanced Resolution Images and Spectrograph) is a second generation system to be installed at VLT-UT4 telescope at Paranal. The instrument is fitted with an embedded Adaptive Optics (AO) module ERIS-AO[2][3][4][5] to drive the adaptive secondary mirror of the UT4, or Deformable Secondary Mirror (DSM)[6][7].

The ERIS-AO module has been integrated at INAF-Osservatorio Astrofisico di Arcetri in Italy and subjected to a year-long verification and qualification test campaign. In order to run the close loop operations with the DSM once on the mountain, a dedicated test tool was developed and is described in this paper.

The DSM Simulator (DMS) is designed to reproduce the light beam from the UT4 Cassegrain and feed the ERIS WFS. The test tool includes a deformable mirror (DM) placed at the system pupil, two light sources for natural and laser guide star, a pupil relay optics producing also the UT4 F/# and an attenuation system to simulate different star magnitudes.

The simulated stars are produced by two optical fibers, chosen with a proper core size to match the desired angular size for natural star (the diffraction limit) and laser star (0.1 to 1.5 arcsec TBC). The fibers are installed on two XYZ stages to allow the precise alignment of the sources within the field. The two beams are mixed by means of a 1% reflective neutral density, used in reflection on the NGS arm and in transmission on the LGS arm, which is monochromatic.

The deformable secondary is the system stop and is simulated by an ALPAO DM, controlled by means of an electronic board (BCU) which is a copy of the DSM one, except the specific drivers for the ALPAO DM. The DM is flat and is placed at a 15° angle wrt the optical axis; the illumination system consists in two fiber launchers (one for LGS, one for NGS) whose beams are mixed by a 1% reflective neutral density, used in transmission on the LGS side. The light cones are collimated by an off-axis parabola to the DM, then focused again by a second off-axis parabola. A spherical mirror picks the beams after the focus and reflects them to a flat folding, to steer the beam toward the instrument dichroic, placed below the DMS bench.

We summarize in the following table the most relevant features of the DMS.

Table 1 The DMS in a nutshell

The spherical mirror is responsible for the final F/# and magnification. Pupil position and field are adjusted by a

| Item | Value | Notes |
| --- | --- | --- |
| Pupil diameter | 21.7 mm | The pupil mask is elliptical to obtain a circular shape when observed at 15°. |
| Magnification | 2.27 | |
| Focal extraction | 203 mm to Cassegrain flange +500 mm to NGS focus +62 mm to LGS focus | Nominal values |
| Output plate scale | 0.00185 arcsec/um 540.54 um/arcsec | |
| NGS wavel. range | 400 nm to 2.8 um | |
| LGS wavel. range | 590 nm ±5 nm | |
| Attenuation range | 0 OD to 6 OD | For both NGS and LGS |
| Fiber input format | FC/PC for NGS SMA for LGS | Same format for the fiber launcher and for the attenuation package. |
| Lamps power supply | 220V | Specific power supply provided for NGS and LGS lamps. |

combined regulation of both spherical mirror and flat folding mirror.

The ALPAO DM used is a DM277, a 277-actuator providing a device with suitable speed and a reasonable number of actuators to allow system verification at PAE.

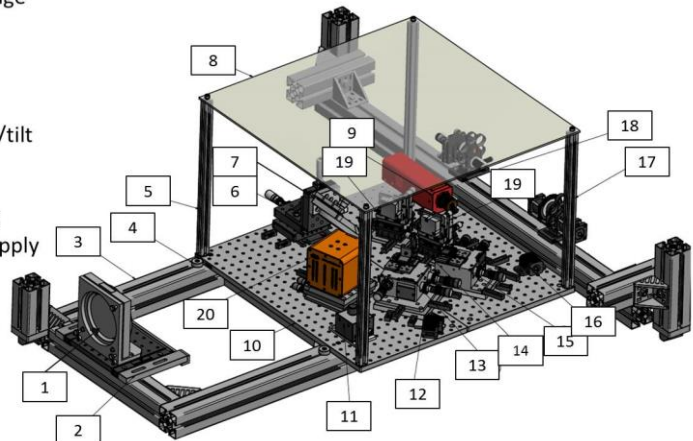

1. Spherical mirror and tip/tilt
2. Spherical mirror stage
3. Rexroth structure
4. Reference pin
5. Roof beam
6. Fold mirror stage
7. Fold mirror and tilt/tilt
8. Protective roof
9. NGS thermal lamp
10. Optical breadboard
11. LGS lamp power supply
12. Test LED lamp
13. Lgs fiber launcher
14. Beam combiner
15. Ngs fiber launcher
16. LGS LED lamp
17. LGS attenuator
18. NGS attenuator
19. OffAxis Par #1, #2
20. Deformable mirror

Fig. 1 Mechanical layout of the DMS with its components installed on the support structure.

# 2 INTEGRATION AND VERIFICATION

## 2.1 Components and assembly

In the following table we give the codes of the DMS components, according to the list in Fig. 1.

Table 2 Components of the DMS

| Item | Description | Notes | Item | Description | Notes |
|---|---|---|---|---|---|
| 1 | Spherical mirror, with tip/tilt mount | | 2 | Spherical mirror stage | For focusing |
| 3 | Support structure, aluminium profiles | | 4 | Reference pin for alignment | |
| 5 | Roof support structure | | 6 | Fold mirror stage | X,Y,Z micrometers |
| 7 | Kinem. holder for the folding mirror | tip/tilt | 8 | Protective roof | |
| 9 | NGS thermal lamp | | 10 | Optical beadboard | 2 x 30x60 cm |
| 11 | LGS LED lamp power supply | Narrow band, 590 peak | 12 | LED Test lamp | |
| 13 | LGS launcher, fiber kinematic. holder | X,Y,Z micrometer | 14 | Beam combiner | 90/10 (T/R) reflective neutral density |
| 15 | LGS launcher, fiber kinematic. holder | X,Y,Z micrometer | 16 | LGS LED lamp | |
| 17 | LGS attenuator | See in the following | 18 | NGS attenuator | See in the following |
| 19 | Off-axis parabola | | 20 | ALPAO-DM277 | |

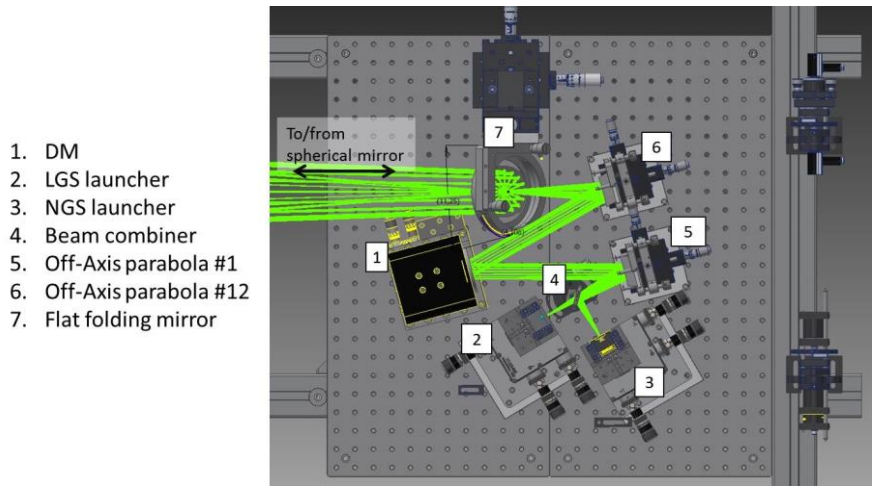

Fig. 2 Optical layout of the DMS main breadboard, showing the light path from the fiber launchers through the beam combiner (#4) to the off-axis parabolas and ALPAO DM.

## 2.2 Internal alignment

One of the driving requirements during the integration and alignment phase was to place the intermediate (or "trapped") focus exactly at the center of the hole in the optical board. The intermediate focus is produced by the parabolic mirror #2, which is fed by a collimated beam, exactly at its focal distance. In order to accomplish such requirement, we developed an integration and alignment procedure, as follows.

1. We installed a fiber launcher at the intermediate focus position, with the goal to materialize that position. In parallel, we installed a fiber launcher at the LGS position. The two fiber cores shall be, after the alignment, each the focal point of the other, thus providing a further check for the alignment.

2. We then install the parabola #2 and a flat mirror (in place of the DM) at their nominal position. We checked the illuminated footprints on both mirrors.
3. We fine-aligned the parabola with a shear interferometer to get rid of coma, focus and astigmatism and deliver a collimated beam to the flat mirror (see detailed procedure below).
4. We placed a pellicle beam splitter in between the launcher and the parabola and measured the return PSF on a high resolution camera.
5. We aligned the flat at 15° wrt the input collimated beam, by using a laser pointer in autocollimation.
6. We installed the parabola #1 at its nominal position.
7. We identified the illuminated footprint on the flat mirror.
8. After shining a beam from the LGS launcher, we adjusted the parabola #1 in tip/tilt to match the footprint on the flat mirror.
9. We then fine aligned the parabola #1 (X,Y) with the shear plate, as already done with the parabola #2
10. Steps 8 and 9 were iterated until the spot from the launcher at the exit focus was focused at the LGS fiber and the beam from parabola #1 to flat mirror was collimated (as from the shear interferometer).

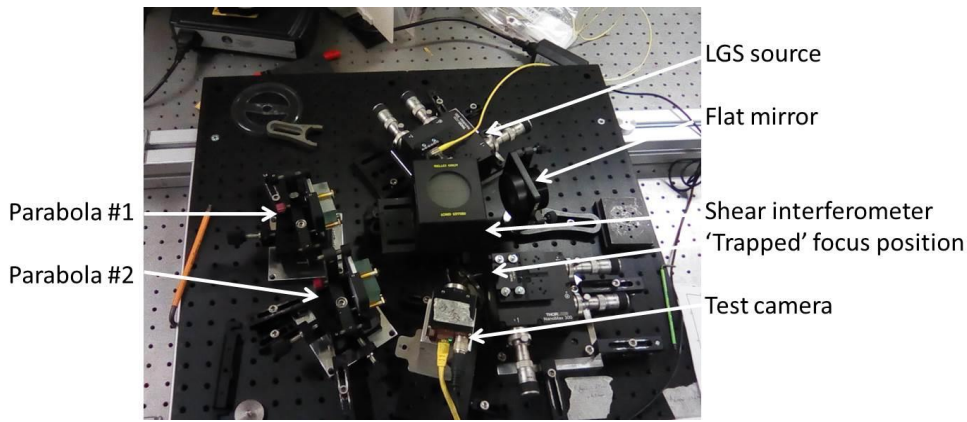

Fig. 3 Main breadboard of the DMS during the alignment and verification of the relay system from the fiber launcher to the trapped focus (materialized by the fiber connector). In the picture are visible the shear plate for the collimation of the first parabola and the camera to inspect the PSF at the trapped focus.

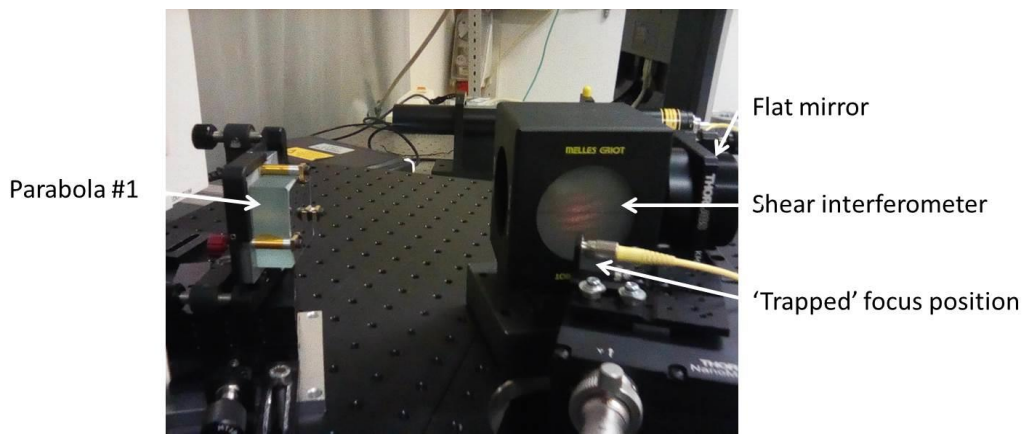

Fig. 4 Shear plate fringes during the alignment of the first off-axis parabola.

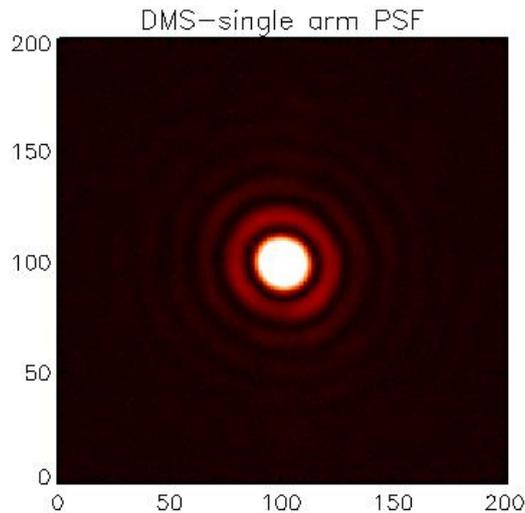

Fig. 5 PSF measured on a single DMS arm (parabola and flat mirror in autocollimation) during the optical alignment.

Thanks to the contemporaneous control of focus and pupil position in both propagation directions, the procedure converged successfully and allowed to have the exit focus exactly at the needed position. We compared of the single-arm PSF with a synthetic one and found it was consistent with a $\lambda/10$ WFE at 632 nm.

After the alignment with a flat mirror, we replaced it with the ALPAO DM, with the following procedure:

1. The spot position produced by the pointer used at step 4 was at the flat mirror center;
2. We removed the flat mirror and installed the DM at its nominal position;
3. We shifted the DM to have the spot at the DM center;
4. We tilted the DM to have the pointer beam reflected back at the pointer itself (DM in autocollimation).

**2.3 PSF inspection**

After the installation of the DM, we verified the optical quality of the system, by measuring the PSF at the exit focus (i.e. after the spherical mirror) with a high resolution camera. By inspecting the PSF, we also optimized the DM flattening command. The result for the NGS side is shown in the following picture, left side: the system is illuminated with a monochromatic source (594 nm) and through a diffraction limited fiber core (P1-630A), whose expected FWHM is approximatively 20 mas, corresponding to 1.5 pixel in the image. The core is saturated on purpose to show the diffraction rings; the saturation caused an evident blooming effect on the image.

The LGS side has been adjusted to obtain a focused PSF at the correct distance wrt the NGS focus (58 mm), including launcher position offset, beam combiner thickness and lateral shift. After adjustment of the differential focus, the LGS source has been laterally offset while monitoring the resulting PSF to obtain an aberrated PSF consistent with the simulated one. The result is shown in the right side of the following picture, where also we compare the actual LGS asterism with those obtained with Zemax in two fields.

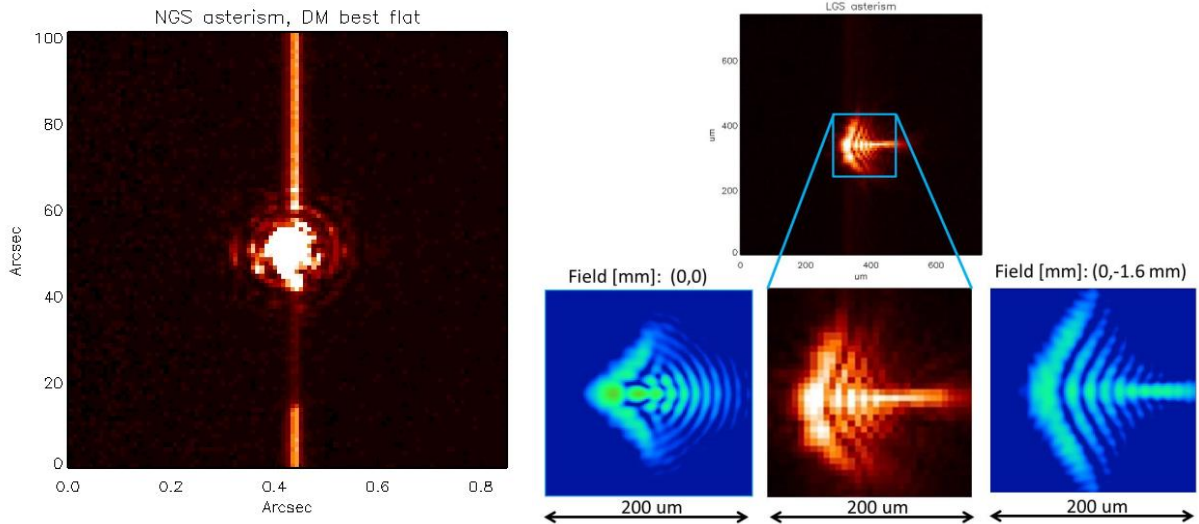

Fig. 6 Left side: NGS diffraction limit source psf, after DM flattening. Right side: LGS PSF measured at different positions in the field.

After adjustment, a best flattening was computed with the DM to calculate the amount of added aberration, estimated as 250 nm Z6 (astigmatism) and 100 nm Z8 (coma). The resulting PSF after correction with the DM is in Fig. 7, right panel.

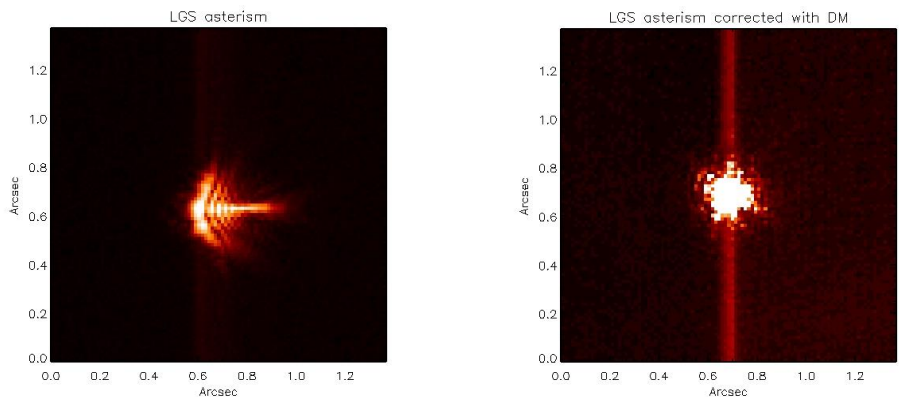

Fig. 7 Left panel: LGS psf, after adding a lateral offset to introduce the LGS off-axis (color scale stretched to enhance the PSf visibility). Right panel: LGS psf after correction with the DM.

### 2.4 Measurement of F-number

The system F-number has been checked after the alignment, by measuring the beam aperture at several positions before and after the focus. To this end, we placed a custom circular mask in front of the DM, in order to produce a perfectly circular shape and avoid ellipticity; the mask has the same size of the final desired aperture. We placed a rail along the exit beam and fine-aligned it with the beam to get rid of projection errors. We measured the out of focus psf on a high resolution camera placed at both intra and extra focal positions (8+8), at 10 mm steps. The F-number has been evaluated by fitting the beam aperture and the measurement positions; additionally, the result has been compared with a fitting algorithm based on the Proper Lib. The results are in the following table and are consistent within 0.4%.

| Mask size | F/# (by fitting) | F/# (Proper Lib) |
|---|---|---|
| 21.7 mm | 13.41 | 13.36 |

# 3 THE ALPAO-DM CALIBRATION

## 3.1 Calibration procedure

The DM calibration procedure is based on interferometric measurements of the mirror surface and a general review is in the references [8]. A specific work on the ALPAO DM277 may also be found in the references[9]. We used a 4D-Technology Twyman-Green interferometer with a beam expander to illuminate the entire active aperture of the DM. The beam expander (see picture below) is composed by two confocal lenses (L1: 25 mm diameter, 90 mm focal length; L2: 40 mm diameter, 400 mm focal length). The lenses have been aligned using a reference flat mirror placed behind them (and previously aligned with the interferometer) and the residual WFE of the beam expander is 20 nm RMS.

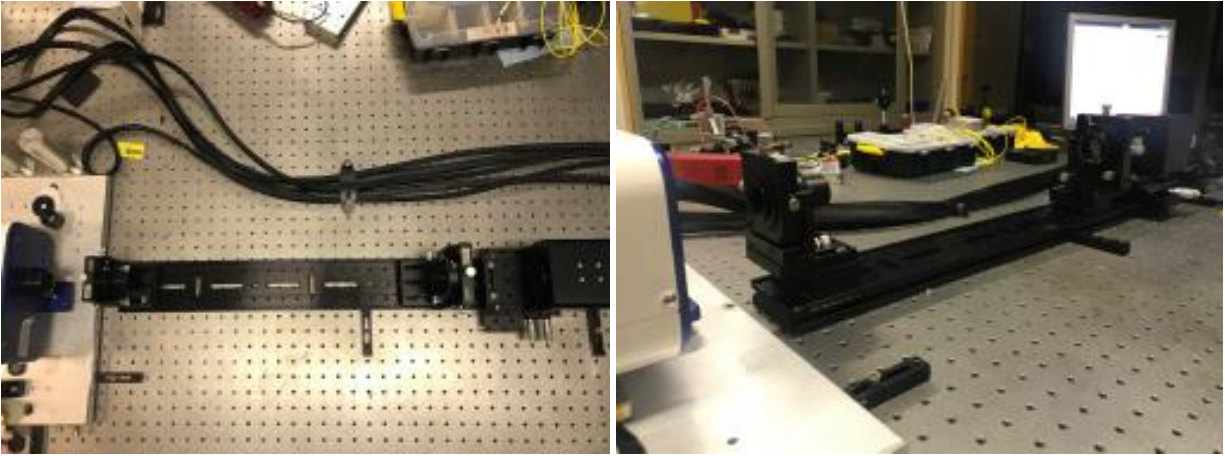

Fig. 8 Optical bench for the calibration of the ALPAO DMs: 4D interferometer (white box on the left) and beam expander on its rail. The DM is mounted on its pitch-yaw stage.

The calibration procedure consists in two steps: at first the actuators are excited and their response is recorded by the interferometer. The dataset is then collected together to create the system interaction matrix M, whose pseudo-inverse R is a projector from the WF space to the actuators one. As a second step, the current WF image w is acquired and the corresponding nulling command c is computed as c= -Rw. The resulting shape is acquired to evaluate the flattening residue, specifically the standard deviation of the pixel-wide WFE after removing the tip-tilt.

In the following figures the starting DM WFE is shown agains the final WF after the flattening procedure. The flattening residue is as low as 12 nm RMS WFE.

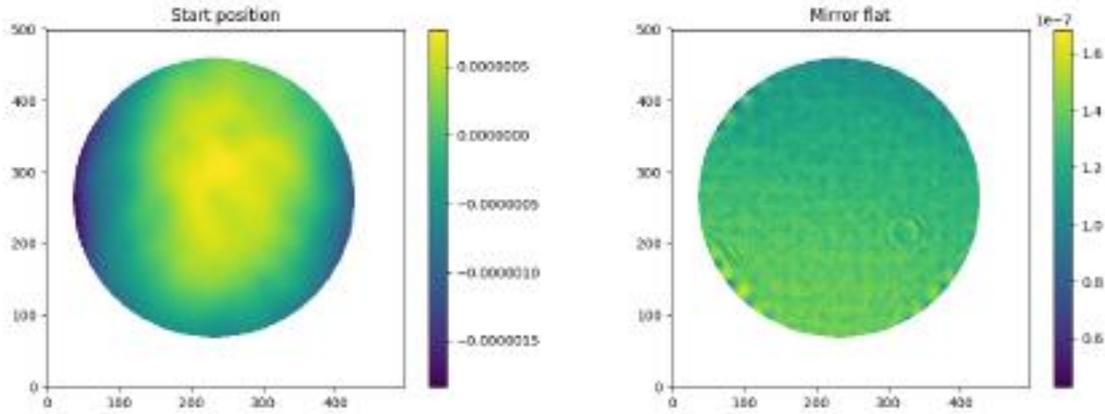

Fig. 9 Comparison between initial and final surface error (left-right) for the DM277.

**3.2 Procedure for the KL matrix fitting**

After the measurement of the actuators influence functions, piled up in the interaction matrix M, we computed a KL command matrix according to a specific procedure. The command matrix, in facts, shall be used with the DM working at a 15° angle of incidence, with a custom elliptical mask to produce the correct pupil shape on the DMS. The KL fitting procedure is as follows:

1. The M matrix is loaded and the median value of each image is subtracted. The median is a robust estimation of the plateau in the frame, representing the rest area of the membrane;
2. Each value is scaled to compensate for the 15° projection;
3. An elliptical mask, with custom dimensions to produce the correct aperture in the DMS, is designed and the pixel within the mask are extracted. The procedure for the mask positioning within the frame is iterative and is based on the identification of the actuators which are visible in the WFS frames.
4. The resulting frame is interpolated to a 240x240 resolution, to match the frame size requested to run the fitting algorithm in SPARTA.

**3.3 Production and verification of the pupil mask**

Since the DM is placed at a 15° angle wrt the optical axis in the DMS layout, we need a custom mask to produce a circular stop with the proper diameter and obtain the F/13.4 aperture. Such a custom mask has been produced according to the following conceptual design: an elliptical obscuring ring is mounted on a conical support, which is interfaced with the DM external envelope. Two main requirements drove the production of the mask: the obscuration size shall be accurate to a fraction of subaperture, i.e. to 0.1 mm, and the safety of the DM optical surface. In facts, the mask with its conical support shall be placed as close as possible to the DM membrane: to guarantee its safety, we performed an accurate verification of the as-built to prevent any possibility for damage while inserting the support.

The mask has been 3D printed and several samples have been produced to identify the best printing procedure. In the end, we found the following production sequence:

1. the printing design is delivered: the minimum distance between the DM front face (envelope) and the mask shall be 4.65 mm (ALPAO communication);
2. the printing process is horizontal;
3. the mask ring is mechanically verified to measure the major and minor axes;
4. the mask is verified on the microHite measuring facility to evaluate the cone support height wrt the interface to DM envelope; the items that doesn't pass the verification are eliminated;
5. the mask ring is smoothed to eliminate any residual astigmatic shape;
6. the mask is verified on the interferometer to measure the best ellipse and to identify any cosmetic defects.

After selection, the mask is inserted in the DM, checking visually for interferences and using an incident light to look at the mask reflection on the mirror. For additional safety 3 shims, 0.8 mm thick, are secured on the DM envelope to act as interface for the mask support. The shims are secured with kapton tape; the mask is secured in position with aluminium tape.

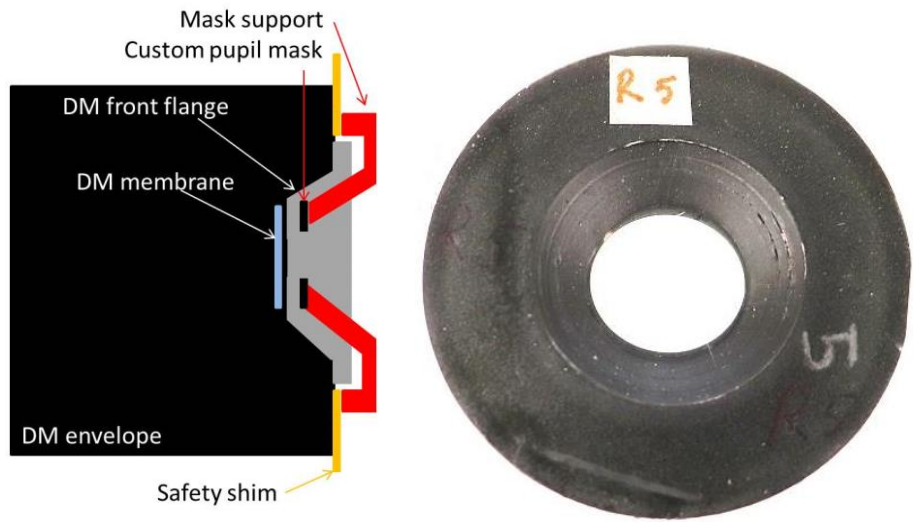

Fig. 10 Left panel: scketch of the pupil mask concept. Righ panel: a sample of pupil mask during the production process.

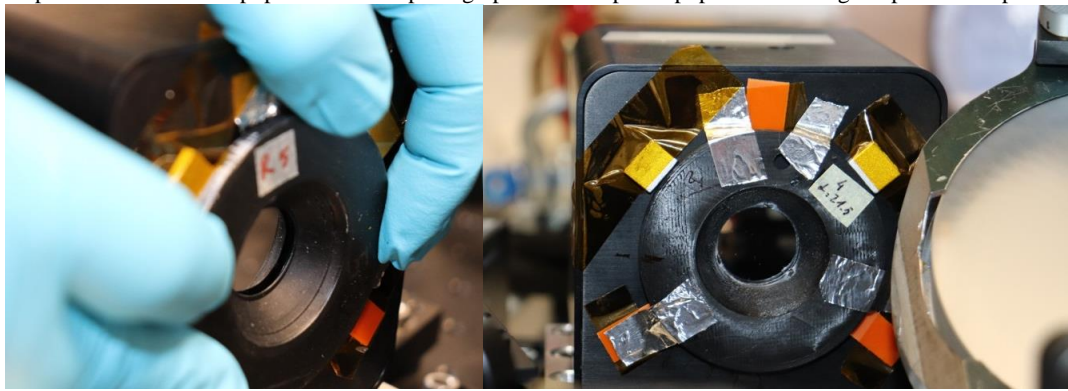

Fig. 11 Left panel: procedure for the mask mounting on the DM envelope. Right panel: the mask after installation. The orange pads are the 0.8 mm shims. The two pads at 120°, covered with yellow kapton tape, are a mechanical reference for the replacement of the pupil mask.

In the following picture we summarize the main features of the pupil mask used.

| Mask Id | Installation date | Mask sizes | Mask distance* | Projected obscuration | Notes |
|---|---|---|---|---|---|
| #4 | 20190827 | 21.5; 22.4 | 0.18+0.8 | ~1 subap larger than desired pupil. | Ok for LGS and HO, not for LO due to light contamination on outer pixels. Final shape is slightly elliptical. |
| #R5 | | 22.1; 20.3 | 0.25+0.8 | ~1 subap smaller than desired pupil. | |

* (wrt DM membrane)

# 4 ARTIFICIAL ASTERISMS

## 4.1 Natural guide star

**NGS Light source**

The lamp selected is a Thorlabs SLS201L, equipped with a tungsten-halogen bulb. The lamp is fitted with an internal feedback system to provide a highly stable power output. The spectral range is 360 nm to 2600 nm, peaked at 1300nm (while the fitted black-body peaks at 1000 nm); the intensity (relative to peak) is 50% at 500 nm (techviewer band), 80% at 800 nm (WFS band) and almost 100% at H-band (test SWIR camera).

The internal optical scheme includes a collimating lens and a NA 0.6 aspheric lens to couple the beam into a fiber. The output flange is equipped with a fiber adapter to connect a multimode, SMA fiber.

Since the lamp it is equipped with a ventilation fan (although low noise), is has been installed outside the DSM optical bench.

**NGS optical fibers**

The NGS side is equipped with the following optical fibers.

| Fiber id | Core size | NA | On sky FWHM | Connector | Notes | Apodization |
|---|---|---|---|---|---|---|
| SMF28E | 9.2 um* | 0.14 | 42 mas | FC/PC-FC/PC | Diff. limit | 0.59, 0.88 |
| M74L02+ | 400 um | 0.39 | 1.7 " | FC/PC-FC/PC | Ext source | na |

*measured at 1310 nm.
+identical (apart from connector, FC vs SMA) to the ext. source one on the LGS side.

The DL fiber has been selected because of its low apodization (measured). The apodization test was performed with the following procedure:

1. A test camera was installed at an extra-focal position, after the exit focus of the system.
2. Several fibers were connected to the lamp and the fiber lancher;
3. For each fiber, an image was taken
4. For each image, an horizontal profile was extracted and smoothed to get rid of the high frequency noise.
5. We evaluated left-side and right-side apodization by comparing the profile value at the end of external diffraction ring with the mean value of the plateau.

The results are shown in the following picture.

The fiber was tested with the SWIR test camera and showed a significant throughput in H-band.

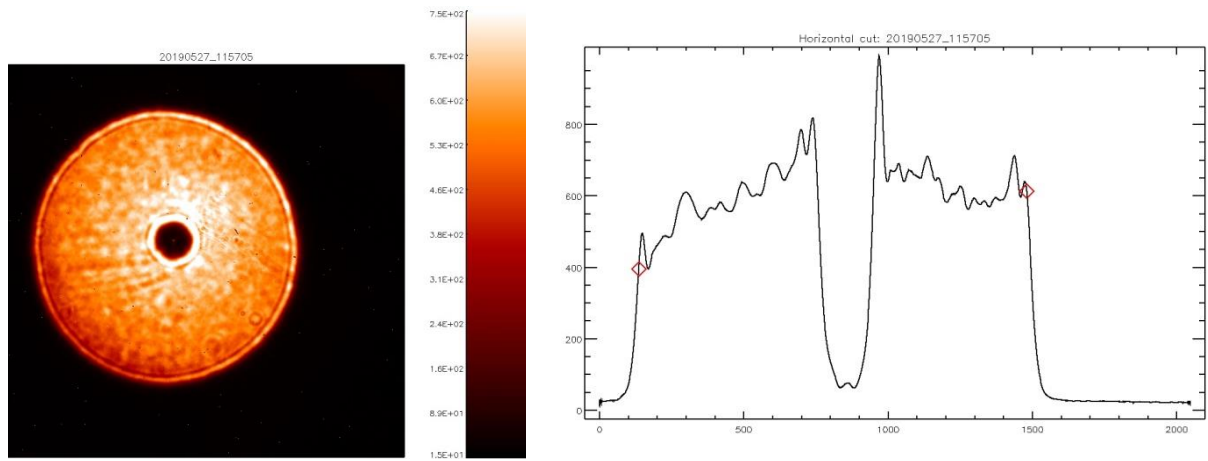

Fig. 12 Out of focus illumination pattern to show and measure the apodization due to the chose fiber.

**NGS photons count**

The NGS side, including thermal lamp and selected optical fiber, has been verified in terms of the light throughput of the system, according to the following procedure:

1. A 600 nm long-pass filter was placed inside the NGS thermal lamp.
2. A visible-band test camera was installed at an extra-focal position after the exit focus; the camera has a negligible QE beyond 800 nm;
3. An image was taken and the total photon count (inside the illuminated area) is measured;
4. The counts have been converted into a photon flux, in the band 600 nm – 800 nm, according to QE in the band and exposure time.
5. The visible-band test camera was replaced by a SWIR test camera.
6. A 1600 nm, 12 nm bandpass was installed on the camera.
7. An image was captured and analysed with the same procedure. The photon count in H band was scaled with the filter bandpass.

As a result, we counted 1.5e10 photons in the range 600-800 nm and we evaluated a lower limit of 3.6e8 photons in H-band.

**4.2 Laser Guide Star**

**LGS Light source**

The LGS source is a LED lamp coupled to an optical fiber (SMA connector). The LED has been preferred (with respect to a 594 nm laser) to avoid speckles using large fiber cores diameter, since the LGS asterism is not diffraction limit. The LED type has been selected for a best match with the laser wavelength, for a minimum leak outside the laser dichroic band (585 nm -595 nm) and for an optimal power output in the central band. The selected type is a Thorlabs M590F2: the peak wavelength is 590 nm with a bandwidth of 16 nm. The lamp comes with a specific power supply providing also power regulation to set the lamp to the desired intensity. In order to minimize the optical power degradation due to increased temperature of the LED junction, the lamp is fitted with a heat sink mounted directly on the LED. From the datasheet, the typical output power (in conjunction with a 400 um core fiber) is 4.6 mW.

**LGS Optical fibers**

In the following table we list the suite of optical fibers used for the LGS asterism. We considered different core sizes to simulate different elongations on sky.

| Fiber id | Core size | NA | On sky FWHM | Connector | Notes |
|---|---|---|---|---|---|
| M14L | 50 um | 0.22 | 0.2 " | SMA-SMA | For alignment and testing |
| M38L01 | 200 um | 0.39 | 0.6 " | SMA-SMA | Nominal elongation |
| M28L01 | 400 um | 0.39 | 1.7" | SMA-SMA | |

**4.3 Attenuation packages and magnitude calibration**

The bench is fitted with a variable attenuation system, in order to feed the WFS with guide stars of different magnitudes. On both sides the attenuation is realized through a 6-position filterwheel, housing 5 metalic, reflective neutral densities (the first position is empty) and a continuous metallic neutral density for fine adjustment of the attenuation. The density is placed within a collimated beam, so that each subsystem is equipped with a collimation/focusing component, namely an aspheric lens for the LGS side (which is highly monochromatic) and a fiber-fed, off axis parabolic collimator for the NGS side. Both subsystems are integrated within a 30 mm cage system. In the following picture the design and as built of the attenuation packages is presented.

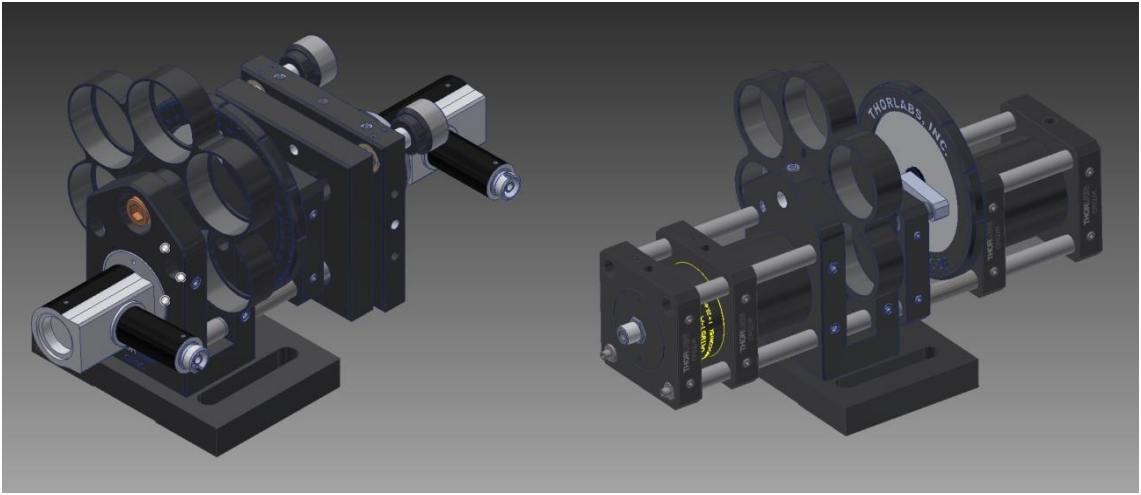

Fig. 13 Design of the NGS (left) and LGS (right) attenuation package.

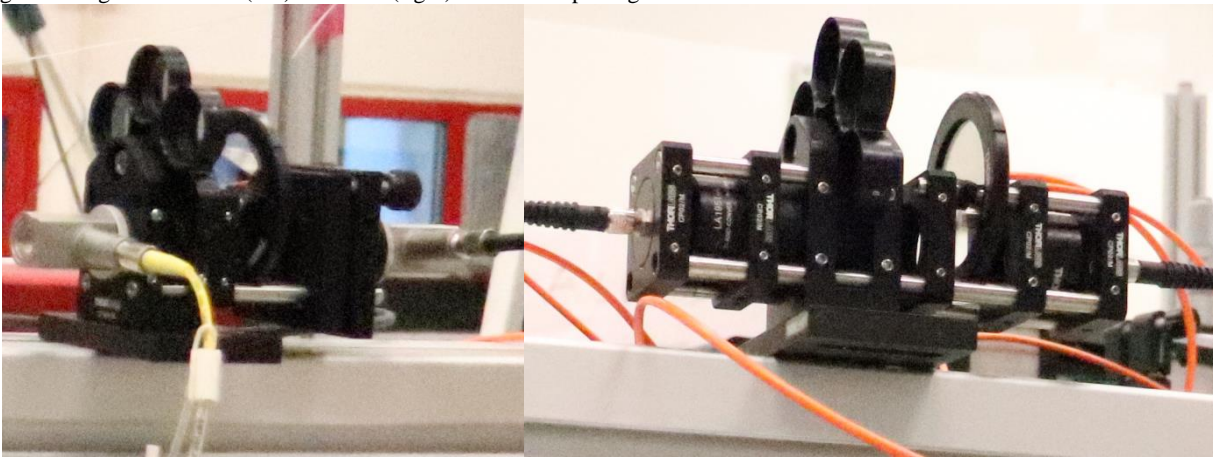

Fig. 14 Attenuation package after integration and installation on the DMS support structure. Left, NGS; right, LGS.

**Neutral density optics**

Each system is equipped with UV fused silica, reflective, nickel coated neutral densities, as in the following table.

| Filter pos | 2 | 3 | 4 | 5 | 6 |
|---|---|---|---|---|---|
| OD nominal | 1 | 1.3 | 2 | 3 | 4 |
| Nominal Attenuation | 10 | 20 | 100 | 1000 | 10000 |

The continuous neutral density is a UV-fused silica NiCrFe coating (on the front face). The density varies in the range 0-2 along a 270° wheel rotation from 90° to 360°, according to the following law (where OD is the optical density and a the angle in degrees)

  *OD=0.00741 *(a-90°)*

**Integration of the LGS attenuator**

The integration of the attenuation package for the LGS side is performed according to the following procedure:

1. The fiber adapters are mounted on the cage plates with a retention ring to set the proper distance.
2. The lenses are mounted on the opposite side of the cage plates.
3. A light is fed through the fiber and the lens distance is adjusted to obtain a collimated beam (distance is regulated with the retention ring on the fiber plate).
4. The 30 mm cage system is built up with the rods and all parts are installed.
5. Filter wheel and variable attenuator are secured to the cage system with the base rods only (the upper ones would interfere with the optical beam).
6. The base plate is bolted below the filterwheel and is used to secure the subsystem on the DSM-Simulator board.

**Integration of the NGS attenuator**

Due to the presence of the reflective fiber collimators, the integration procedure of the attenuation package for the NGS side is different from the LGS one and is described below.

1. The entrance collimator is installed on the SM1 tube with the adapter ring and then attached to the tip-tilt mount.
2. The 30 mm cage system is built up with the rods and all parts are installed.
3. Filter wheel and variable attenuator are secured to the cage system with the base rods only (the upper ones would interfere with the optical beam).
4. The base plate is bolted below the filterwheel and is used to secure the subsystem on the DSM-Simulator board.
5. The exit collimator is installed on the filterwheel with the adapter ring.
6. A light is fed through the entrance fiber and the tip-tilt is adjusted to maximize the light intensity on the exit fiber.

**Neutral density verification**

The ND filters have been qualitatively checked upon arrival to rule out manufacturing errors or packaging mismatches. The measured densities were consistent with the specifications. After integrating the attenuation device, the NGS system was characterized. To this end, we measured the light flux at all the discrete filterwheel positions and at 5 angles of the continuous density filter. The light flux was measured by a CCD installed in place of the exit parabola, as the total counts over the illuminated area, after subtracting the dark current associated with the given integration time. The counts have been then normalized to 1 s integration time and compared with the no-filter count. The light source is the thermal lamp SLS201, coupled with the attenuation package via a M18 fiber (100 um core) to avoid saturation at the bright end.

The results for the discrete filter wheel are shown in the following table. The measured values are consistent with the nominal attenuation, with an excess of uncertainty at the faint end. These data are intended for qualification of the system, while the final filter-to-magnitude calibration will be fine-tuned with the WFS.

| **Filter pos** | **1** | **2** | **3** | **4** | **5** | **6** |
|---|---|---|---|---|---|---|
| OD nominal | 0 (no-filter) | 1 | 1.3 | 2 | 3 | 4 |
| Attenuation nominal | 1 | 10 | 20 | 100 | 1000 | 10000 |
| Measured attenuation | 1 | 11.7 | 20.5 | 94.5 | 623 | 6920 |

In the next table we present the attenuation (measured and nominal) for the continuous density at 5 positions. 30° corresponds to "transparent window". The experimental values are consistent with the nominal attenuation, also considering the uncertainty in the wheel positioning.

| **Wheel position** | **30°** | **120°** | **210°** | **300°** | **340°** |
|---|---|---|---|---|---|
| OD nominal | 0 | 0.22 | 0.89 | 1.55 | 1.85 |

| Attenuation nominal | 1 | 1.67 | 7.75 | 35.9 | 71.2 |
|---|---|---|---|---|---|
| Measured attenuation | 1 | 1.89 | 9.1 | 35.7 | 76.9 |

## 5  INTEGRATION AND ALIGNMENT WITH THE ERIS-AO

In this section we give a summary of the integration of the DMS with the ERIS-AO instrument and of the alignment procedure. The sequence is specifically defined for the ERIS-AO module; however it can be adapted to different system, thus enabling the DMS to become a testing tool for the AOF.

### 5.1 Mounting procedure

The DMS bench sits on the ERIS-Cassegrain flange by means of three V-grooves to ensure a repeatable mount point and preserve the alignment. On the DMS side, the V-grooves shall be adjusted in order to have the intermediate foci (those laying at the flat mirror central hole) coincident with the spot from the LGS pointer.

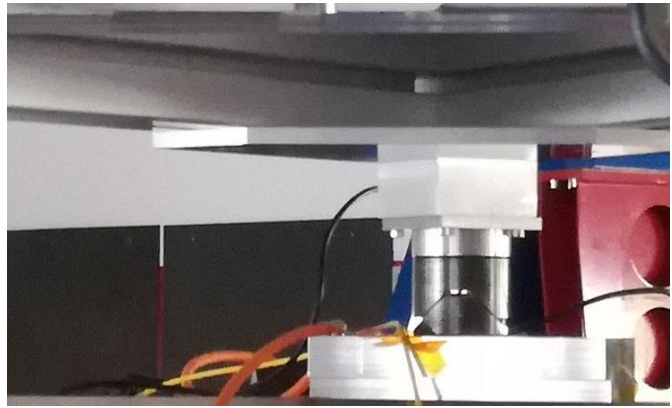

Fig. 15 V-grooves for interfacing the DMS bench woth the ERIS cassegrain flange.

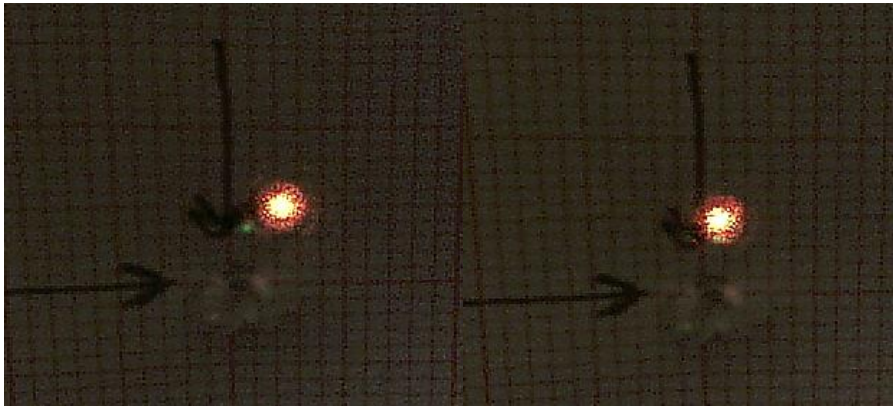

Fig. 16 Position check for the LGS pointer (yellow spot) and the DMS-NGS intermediate focus at the flat mirror hole. The paper sheet is installed on the front surface of the flat mirror and secured to its surface with a nylon wire, to avoid projection effects in the measurement.

### 5.2 Alignment procedure

The optical alignment of the DMS with the ERIS AO includes five steps:

1. The nulling of the focus.
2. The alignment of the pupil position.
3. The nulling of the residual tilt.

4. The centring of the footprint of the fold mirror hole with the pupil.
5. The final, fine nulling of the residual tilt.

Given the optical design, the pupil position and the tilt are controlled by both mirrors (the sphere and the flat folding); however, the pupil position is mostly affected by the sphere angle, while the flat mirror controls the tip-tilt with a much lower effect on the pupil position. The focus is controlled only via the sphere stage; acting on this stage has a significant impact on the tilt, this is the reason why this step is the first in the procedure.

In the following table we list, for each step, the optical element to be acted and the feedback to be used.

| # | Target | Action | Feedback | Notes |
|---|--------|--------|----------|-------|
| 1 | Focus nulled | • Disengage the brake on the sphere stage;<br>• adjust the focus;<br>• set the brake. | Zernike coefficients | Moving the stage, a significant tilt may be added. Be sure that the Zernike reading is meaningfull. |
| 2 | Pupil centered | • act on sphere Rx, Ry<br>• compensate the tilt with the flat mirror Rx, Ry | Intensity panel for the pupil; spot position on the tech viewer for the tilt | |
| 3 | Tilt nulled | • Act on the flat mirror Rx, Ry | Zernike coefficients. | Preliminary correction: it is enough to null the tilt to, say, 100 nm. |
| 4 | Flat mirror hole centered on the pupil image | • Act on the flat mirror stage X,Y to center the hole footprint;<br>• Compensate the tilt with flat mirror stage Z | Subapertures panel for the hole footprint; spot position on the techviewer for the tilt. | |
| 5 | Tilt nulled | • Act on the flat mirror Rx, Ry | Zernike coefficients. | |

In the following table we show the interaction matrices of the various degrees of freedom. Positive means clockwise rotation, i.e. screwing; negative means counter-clockwise, i.e. un-screwing. The pupil-rerotator angle shall be set to a fixed, known position in order to preserve the directions as in the table.

| Item | Effect | Notes |
|------|--------|-------|
| Sphere RX + | Pupil moves up | PSF moves right in techviewer |
| Sphere RY + | Pupil moves right | PSF moves up in techviewer |
| Flat Rx + | PSF moves left in techviewer | |
| Flat Ry + | PSF moves down in techviewer | |
| Flat stage Y + | Hole footprint moves left | In-plane movement, no tilt added |
| Flat stage X + | Hole footprint moves down | Tilt added, to be compensated with stage Z + |